\documentclass[epj]{svjour}
\usepackage{graphicx}
\usepackage{amstext}
\usepackage{array}
\usepackage{amssymb}
\usepackage{amsmath}
\newcommand{\be}{\begin{equation}}
\newcommand{\ee}{\end{equation}}
\newcommand{\bea}{\begin{eqnarray}}
\newcommand{\eea}{\end{eqnarray}}
\newcommand{\Eq}[1]{Eq.\,(\ref{#1})}% \Eq{abc}
\newcommand{\Fig}[1]{Fig.\,\ref{#1}}% \Fig{fig:abc}
\newcommand{\Sec}[1]{Sec.\,\ref{#1}}% \Sec{sec:abc} sic!byc konsekewntnym \label{sec:xx} \Sec{sec:xx}
 \usepackage{epstopdf}
\begin{document}
\title{Resonant states in double and triple quantum wells}
\author{A.\,Tanimu \and E.\,A. Muljarov% etc
%\thanks{\emph{E-mail:} egor.muljarov@astro.cf.ac.uk}%
}                     % Do not remove
\institute{School of Physics and Astronomy, Cardiff University, Cardiff CF24 3AA,
United Kingdom }
\date{Received: date / Revised version: \today}
\abstract{
The full set of resonant states in double and triple quantum well/barrier structures is investigated. This includes bound, anti-bound and normal resonant states which are all eigensolutions of Schr\"odinger's equation with generalized outgoing wave boundary conditions. The transformation of resonant states and their transitions between different subgroups as well as the role of each subgroup in observables, such as the quantum transmission, is analyzed. The quantum well potentials are modeled by Dirac delta functions; therefore, as part of this study, the well known problem of bound states in delta-like potentials is also revisited.
 % end of PACS codes
} %end of abstract
\maketitle
\section{Introduction}
\label{intro}

Resonant states (RSs) have been known in quantum mechanics for almost a century, since the pioneering works of Gamow~\cite{GamowZP1928} and Siegert~\cite{SiegertPR39}. They describe, in a mathematically rigorous way, natural resonances which quantum systems exhibit. People are dealing with resonances in different fields of physics, ranging from classical mechanics and electrodynamics to quantum physics and gravity. Resonant phenomena have attracted significant interest in recent years, in particular, in quantum mechanics due to a rapid progress in the field of semiconductor nanostructures, where different electronic states are formed in various types of quantum potentials. In spite of this growing interest in resonances, many fundamental aspects of RSs in quantum systems are still to be investigated~\cite{HatanoPTP07}.

Perhaps, a more traditional way of dealing with resonances is to study the singularities of the scattering matrix~\cite{Nuss59} as also described in many textbooks (see, e.g.~\cite{MandleBook92}). Finding these singularities is actually equivalent to solving the Schr\"{o}dinger equation with outgoing wave boundary conditions outside the system. However, these boundary conditions strictly define RSs. In general, RSs have complex energy eigenvalues, showing that the states decay exponentially in time, leaking out of the system (such as a quantum well). Early studies of RSs~\cite{MorePRA71,MorePRA73} revealed that they can form a complete set of functions inside the quantum system, and therefore can be used as a basis for expansion, in order to find RSs of a modified system. This idea, first suggested in nuclear physics~\cite{BangNPA1978} has been recently developed in electromagnetics into a powerful method  called resonant-state expansion (RSE)~\cite{MuljarovEPL10,DoostPRA14}.
The RSE uses as a basis the RSs of a simple system, usually analytically solvable. The advantage of applying the RSE to various systems becomes obvious in case of perturbations
which cannot be treated analytically.

The aim of this paper is to study the RSs of simple one-dimensional (1D) quantum-mechanical systems, such as double and triple quantum wells, for better understanding of their properties, as well as for generating an analytic basis of RSs for its further use in the RSE treating more complicated potentials. In this work, we take a well-known simplification of a multiple-quantum well/barrier potential, approximating it with a sequence of Dirac delta functions. Bound states  in such potentials are known the literature~\cite{GriffithsBook05}, as well as the periodic solutions of the famous Kronig-Penney potential~\cite{KronigPenney31} modeling the electronic band structure of a 1D crystal lattice. However, the spectral properties of quantum systems are not limited to bound states. Rather, phenomena, such as quantum tunneling through barriers and quantum transmission and scattering of particles across the potential, are mainly determined by the internal resonances of the system, which are described by the RSs. These, however, to the best of our knowledge, have not been investigated so far even in such simple systems as 1D double and triple Dirac quantum wells or barriers. The present work  is a thorough study of RSs in such potentials.

In this work, we investigate the full spectrum of eigensolutions of the 1D Schr\"odinger equation for double and triple quantum well/barrier systems. The full spectrum of RSs includes bound, anti-bound and normal RSs, all together forming a complete set of functions and determining the spectral properties of a quantum system, such as the local density of states and transmission~\cite{ArmitagePRA14}. We first revisit the bound state problem in double and triple quantum well systems, working out exact solutions and some important asymptotics allowing explicit analytic expressions. Then we demonstrate how the bound states appear or disappear in the spectrum transforming into anti-bound states as the parameters of the potential change. Then we extend our consideration to the full spectrum of RSs and discuss the physical meaning of the normal RSs, also paying attention to the their evolution and transformation into/from bound and anti-bound states~\cite{Nuss59,SprungAJP96,ZavinJPA04,BelchevCJP11}. Finally, we investigate the role of the RSs in the quantum transmission.

\section{Resonant states of one-dimensional quantum systems }

In general, RSs of a quantum-mechanical system are eigen solutions of the Schr\"{o}\-dinger equation
\be
\hat{H}({\bf r})\psi_n({\bf r})=E_n\psi_n({\bf r})\,,
\label{SE3D}
\ee
satisfying the outgoing wave boundary conditions (BCs). Here $\hat{H}({\bf r})$ is the Hamiltonian of a single particle, $\psi_n({\bf r})$ and $E_n$ are, respectively, its eigen wave function and eigen energy, and ${\bf r}$ is a three-dimensional coordinate.  Having in mind application to e.g. planar semiconductor heterostructures, we reduce our consideration in this work to a non-relativistic 1D Schr\"{o}dinger's problem. For bre\-vity of notations, we make use of the units $\hbar=1$ and $m=1/2$, where $m$ is the particle mass (e.g. the electron effective mass in a semiconductor). It is also convenient to introduce the eigen wave number $k_n$ of the particle associated with a given RS and use it instead of the energy $E_n$ which is linked to it via the non-relativistic parabolic dispersion relation
\be
E_n=k_n^2\,.
\label{dispersion}
\ee
A 1D time-independent Schr\"{o}dinger equation then takes the form:
\be
\left[-\frac{d^2}{dx^2}+V(x)\right]\psi_n(x)=k_n^2\psi_n(x)\,,
\label{SE}
\ee
where $V(x)$ is the potential of the particle, which is chosen in such a way that it vanishes outside the system.

In 1D, the outgoing wave BCs for RSs reduce to
\be
\psi_n(x)\propto e^{ik_n|x|} {\rm \ \ for\ \ } |x|\rightarrow\infty\,,
\label{BCs}
\ee
which are also known as Siegert BCs~\cite{SiegertPR39}. Solving \Eq{SE} with the BCs \Eq{BCs} inevitably leads to the fact that the energies $E_n$ are generally complex,
\be
E_n=(p_n+i\varkappa_n)^2=(p_n^2-\varkappa_n^2)+2ip_n\varkappa_n\,,
\ee
where $p_n$ and $\varkappa_n$  are, respectively, the real and the imaginary part of the eigen wave number: $k_n=p_n+i\varkappa_n$.
For bound states $p_n=0$ and $\varkappa_n>0$, so that the energy is real negative $E_n=-\varkappa_n^2<0$, and the general \Eq{BCs} reduces to the standard BC of the wave function vanishing away from the system: $\psi_n(x)\propto e^{-\varkappa_n|x|}\to0$ at $|x|\rightarrow\infty$. For anti-bound states~\cite{ZavinJPA04}, if they exist in the spectrum, $p_n=0$ and $\varkappa_n<0$, corresponding to a purely growing wave outside the system, even though their energies are real and negative.  All other RSs have $p_n\neq0$ and $\varkappa_n<0$ which results in complex eigen energies and wave functions which oscillate and grow exponentially in the exterior: $\psi_n(x)\propto e^{(ip_n-\varkappa_n)|x|}\to\infty $, according to \Eq{BCs}.

As a consequence of this exponential growth, the wave functions of the RSs are not orthogonal and not normalizable in the usual way. RSs instead require a proper general orthonormality condition which would include the standard one as a special case, valid for bound states. For a one-dimensional system, this general orthonormality of RSs is given~\cite{SiegertPR39,MorePRA71,MuljarovEPL10} by
\bea
\delta_{nm}&=& \int_{x_L}^{x_R}\psi_n(x)\psi_m(x)dx \nonumber\\
&&-\frac{{\psi_n(x_L)\psi_m(x_L)}+{\psi_n(x_R)\psi_m(x_R)}}{i(k_n+k_m)},
\label{norm}
\eea
where $\delta_{nm}$ is the Kronecker delta, and $x_L$ and $x_R$ are two arbitrary points outside the system, one to the left of it ($x_L$) and one to the right ($x_R$).
For bound states, it can be easily seen, by taking the limits $x_{R,L}\to\pm\infty $ and noting that the second term vanishes due to the vanishing wave function, that \Eq{norm} leads to the standard orthonormality: $\delta_{nm}=\int_{-\infty}^{\infty}\psi_n(x)\psi_m(x) dx$. For exponentially growing wave functions the divergence of the integral at $|x_{R,L}|\to \infty$ is exactly compensated by the second term.  Furthermore, as the normalization does not depend on $x_L$ and $x_R$, it is usually convenient to take these points exactly at the boundaries of the system.

\section{Double well}
\label{sec:double}

\begin{figure}[t]
\hskip1.3cm
\includegraphics[scale=0.28,angle=-90]{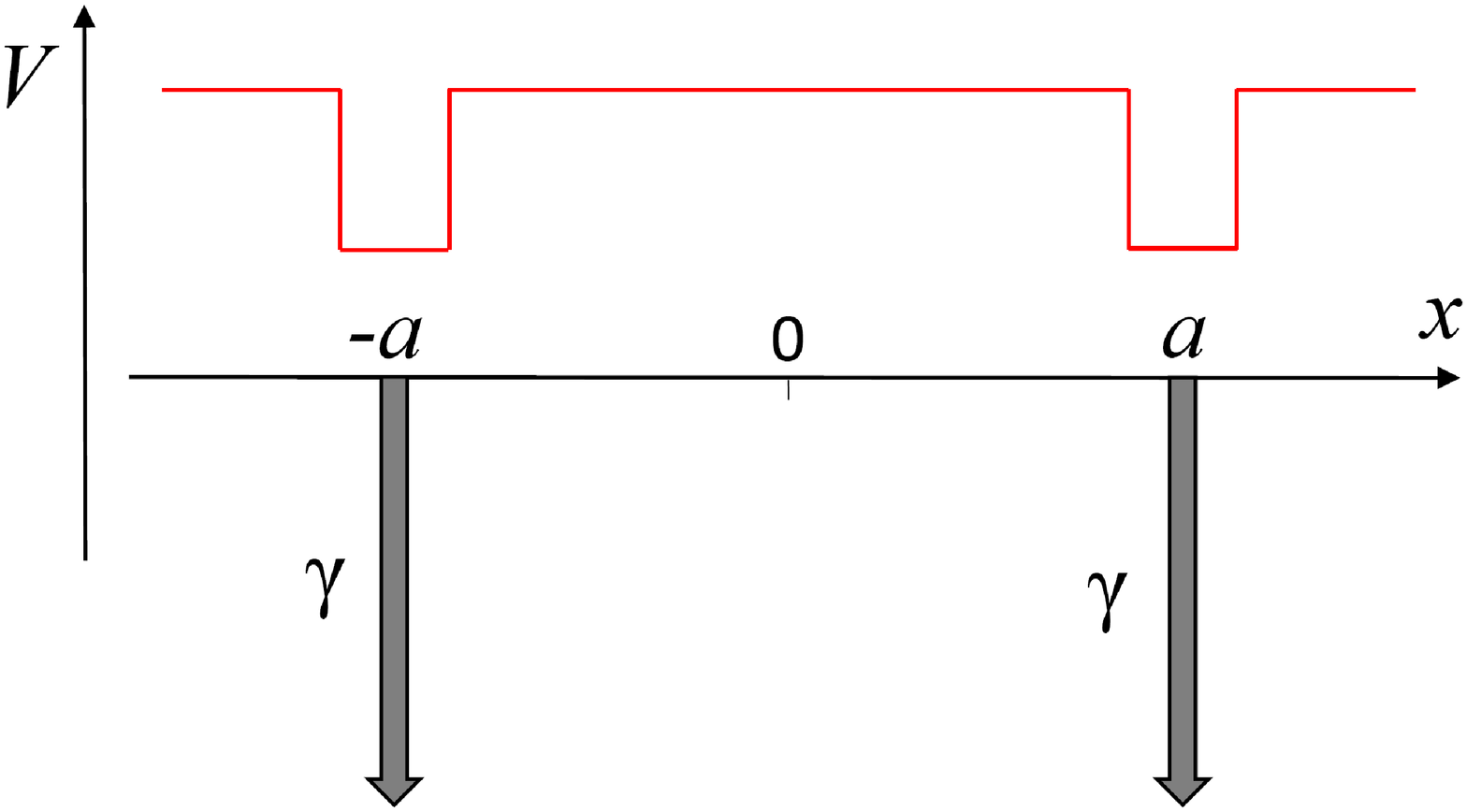}
\vskip-2.5cm
\caption{A sketch of the potential of a symmetric double well structure (red line) approximated by two delta-functions (grey arrows).}
\label{fig:Sketch-double}
\end{figure}

We model a symmetric double quantum well structure by a superposition of two Dirac delta functions,
\be
V(x)=-\gamma\delta(x-a)-\gamma\delta(x+a)\,,
\label{pot-double}
\ee
where $2a$ is the distance between the wells and $\gamma$ is the strength of the potential which has the meaning of the depth of each quantum well multiplied by its width.
Figure~\ref{fig:Sketch-double} sketches this potential along with a realistic coupled quantum well structure it models.
While potentials modeled by delta functions sometimes fail to catch interesting physical phenomena, such as the band crossing~\cite{AllenPR53},
an obvious advantage of this model is its simplicity and explicit analytical solvability. The solution for this potential, in terms of bound states, has been covered in depth in many texbooks, see e.g.~\cite{GriffithsBook05}. The first few resonant states in double barrier structures ($\gamma<0$) were found in~\cite{HatanoPTP07}.
We revisit this problem again, in order to study the full spectrum of RSs and their properties, which has not been done in the literature. This is also of practical importance, as the full set of RSs can be further used as a basis for the RSE.

\subsection{Exact solution}

A general solution of \Eq{SE} with the potential \Eq{pot-double} has the form (for brevity of notations, we drop in this and the following section the index $n$ labeling RSs):
\be
\psi(x) = \begin{cases}
A e^{ikx}+B e^{-ikx}                               &   x>a, \\
C_1e^{ikx}+ C_2e^{-ikx}           &   |x|\leqslant a,\\
D e^{ikx} +F e^{-ikx}                               &   x<-a,\\
\end{cases}
\label{psi0}
\ee
with constant coefficients standing at the exponentials. Applying the outgoing wave BCs leads to $B=D=0$. Furthermore, using the mirror symmetry of the potential, $V(-x)=V(x)$,
splits all the solutions into two groups: even and odd states, having the property
\be
\psi(-x)=\pm\psi(x)\,.
\label{even-odd}
\ee
From this we obtain $F=\pm A$ and $C_1=\pm C_2=C$. Then the wave function takes the form
\be
\psi(x) = \begin{cases}
 A e^{ik x}                               &   x>a, \\
 C\left(e^{ik x}\pm e^{-ik x}\right)          &  |x|\leqslant a,\\
\pm A e^{-ik x}                               &   x<-a.\\
\end{cases}
\label{wf-double}
\ee
The wave function $\psi(x)$ must be continuous at any point but its derivative $\psi'(x)$ is discontinuous at $x=\pm a$.
The break in the derivative can be evaluated by integrating \Eq{SE} across the delta-function potential wells. This yields four boundary conditions determining the relation between the coefficients $A$ and $C$, as well as the eigenvalues $k$. However, as the symmetry of the potential has been already taken into account leading to \Eq{wf-double}, only one pair of BCs (e.g. at $x=a$) provides a unique information:
\bea
\psi'(a+0_+)-\psi'(a-0_+)&=&-\gamma\psi(a)\,,
\label{BC1}
\\
\psi(a+0_+)-\psi(a-0_+)&=&0\,,
\label{BC2}
\eea
where $0_+$ is a  positive infinitesimal. The other pair of BCs (at $x=-a$) is then fulfilled automatically.
Substituting the wave function \Eq{wf-double} into the BCs Eqs.\,(\ref{BC1}) and (\ref{BC2}), obtain
\bea
ik A e^{ik a}-ik C (e^{ika}\mp e^{-ika})&=&-\gamma A e^{ik a}\,,
\label{BC1a}
\\
A e^{ik a}-C (e^{ika}\pm e^{-ika})&=&0\,,
\label{BC2a}
\eea
Expressing the ratio $A/C$ from Eqs.\,(\ref{BC1}) and (\ref{BC2}) and combining the results obtain
\be
\frac{A}{C}=\frac{ik (e^{ika}\mp e^{-ika})}{(ik+\gamma)e^{ika}}=\frac{e^{ika}\pm e^{-ika}}{e^{ika}}\,,
\ee
After rearrangement this yields a transcendental secular equation
\be
1+\frac{2ik}{\gamma}=\mp e^{2ika}
\label{sol-double}
\ee
determining all the RS eigenvalues $k_n$. Note that the upper (lower) sign corresponds to even (odd) RSs.

\subsection{Bound and anti-bound states}
\label{sec:double-bound}

To find bound and anti-bound states of the system, we make a substitution $k=i\varkappa$ in \Eq{sol-double} and solve the latter for real values of $\varkappa$. Then the eigen energy $E=-\varkappa^2$ takes real negative values. For bound states, $\varkappa$ should be positive, as required by the evanescent form of the wave function outside the system. For anti-bound states instead the wave function has a pure exponential growth to the exterior which is provided by $\varkappa<0$.

\begin{figure}[t]
\includegraphics[scale=0.35,angle=-90]{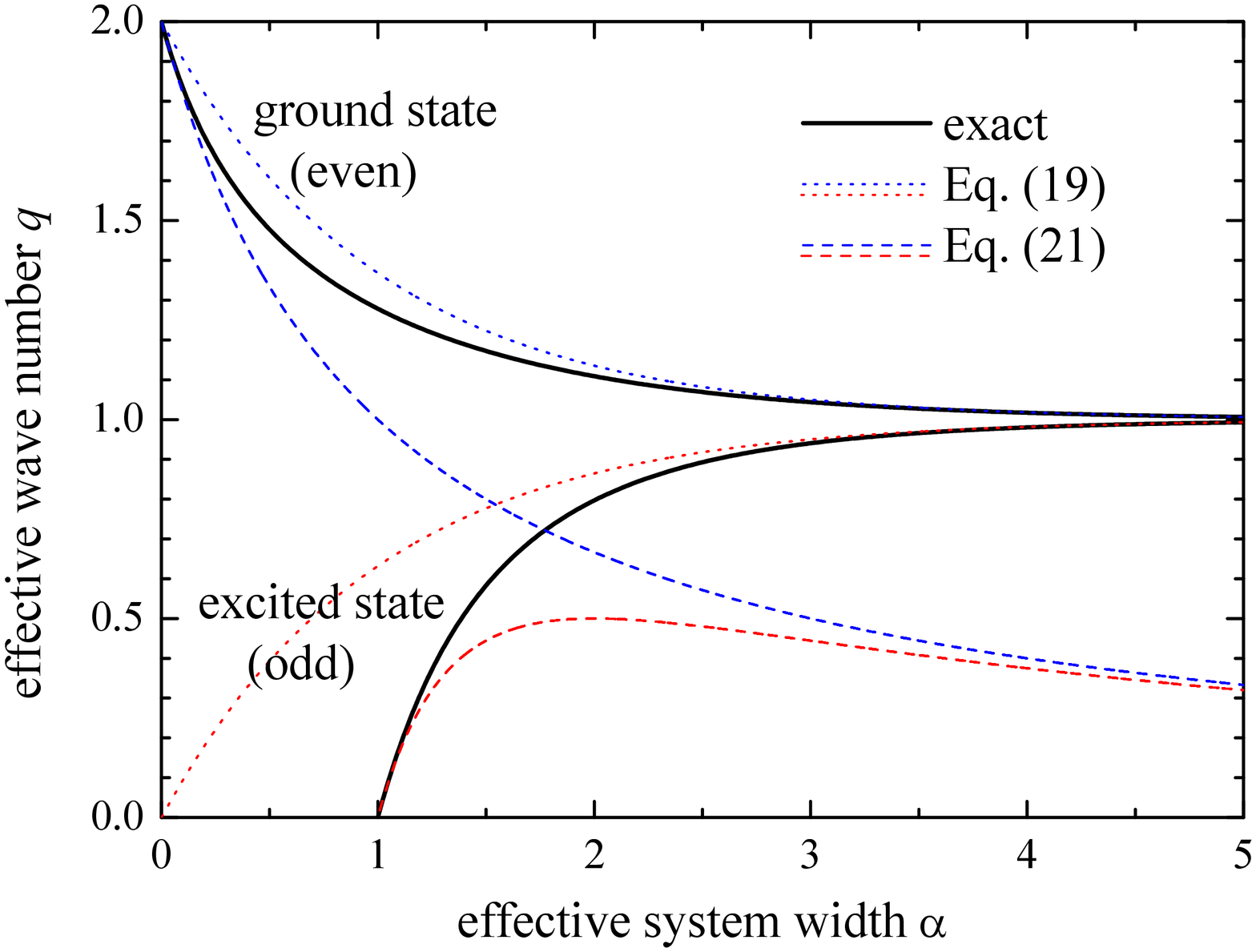}
\vskip-2cm
\caption{Exact (black solid lines) and approximate (blue and red dashed and dotted lines) effective wave numbers $q=-2ik/\gamma=2\varkappa/\gamma$ of the two bound states in a double quantum well system modeled by a double delta function  potential, as functions of the effective width of the system $\alpha=\gamma a$.}
\label{fig:Double-q}
\end{figure}

While the secular equation \Eq{sol-double} apparently depends on two parameters, $\gamma$ and $a$, this parametric space reduces to a single parameter
\be
\alpha=\gamma a
\label{alpha}
\ee
which can be treated as the effective system size or the effective strengths of the potential. Concentrating on the dependence of the eigen states on the system size (e.g. keeping the strength $\gamma$ fixed), it is convenient to introduce a dimensionless wave number $q=2\varkappa/\gamma$. Then \Eq{sol-double} takes the form
\be
q_\pm=1\pm e^{-q_\pm\alpha}\,,
\label{sol-double-q}
\ee
where index $+$ ($-$) labels even (odd) parity states.
The full solution of \Eq{sol-double-q} found numerically with the help of the Newton-Raphson method implemented in MATLAB is shown in \Fig{fig:Double-q} for positive values of $\alpha$ and $q$. It demonstrates the dependence of the imaginary wave vector for two bound (even and odd) states in the system as function of the effective width $\alpha$.

At large distances between the wells ($\alpha\gg1$) the two states are quasi-degenerate,
\be
q_\pm\approx 1\pm e^{-\alpha}\,,
\label{large-q_appr}
\ee
illustrating the fact that each isolated delta-like quantum well accommodates only one bound state with $q=1$. As clear from \Fig{fig:Double-q},  \Eq{large-q_appr} is a good approximation of the full solution \Eq{sol-double-q} for $\alpha> 3$. As $\alpha$ increases, the splitting between the levels becomes exponentially small, reflecting the vanishing tunnel coupling between the wells.

In the opposite limit of small width $a$ or small wave number $\varkappa$ (i.e. small binding energy), one can obtain a simple analytic approximation, based on the Taylor expansion of the exponential in \Eq{sol-double-q},
\be
e^{-q \alpha}\approx 1-q\alpha +q^2\alpha^2/2\,,
\label{Taylor}
\ee
valid for $|q\alpha|=| \varkappa a|\ll 1$.
For the even parity state, it is sufficient to use the expansion \Eq{Taylor} up to 1st order, while the same level of approximation for the odd parity state requires also the 2nd order to be taken into account. Then approximate solutions of \Eq{sol-double-q} take the form:
\be
q_+\approx \frac{2}{\alpha+1}\ {\rm (even),}\ \ \ q_-\approx 2\frac{\alpha-1}{\alpha^2}\ {\rm (odd)}.
\label{double-approx}
\ee
They are shown in \Fig{fig:Double-q} by dashed lines matching the exact solution (solid lines) at small $\alpha$ (for the even state) or at small $q$ (for the odd state).

The analytic approximation \Eq{double-approx} also allows us to find a condition for bound states to exist in the system, which requires that $q>0$. Indeed, when a bound state disappears from the spectrum, its binding energy vanishes, meaning that $q\to0$. This makes the approximation \Eq{double-approx} valid, so that it precisely determines the critical values of the system parameters when the bound state disappears. While the ground state exists for any $\alpha>0$ ($q_+$ is always positive), the excited (odd) bound state exists only for
\be
\alpha>1
\label{odd-inequ}
\ee
and disappears at $\alpha=1$ (when $q_-$ is vanishing), as the width of the system becomes insufficient to accommodate it, given the tunnel coupling between the wells. However, a quantum state itself cannot disappear from the system completely. Instead, it transforms into an anti-bound state which can be observed for $\alpha<1$.

\begin{figure}[t]
\includegraphics[scale=0.35,angle=-90]{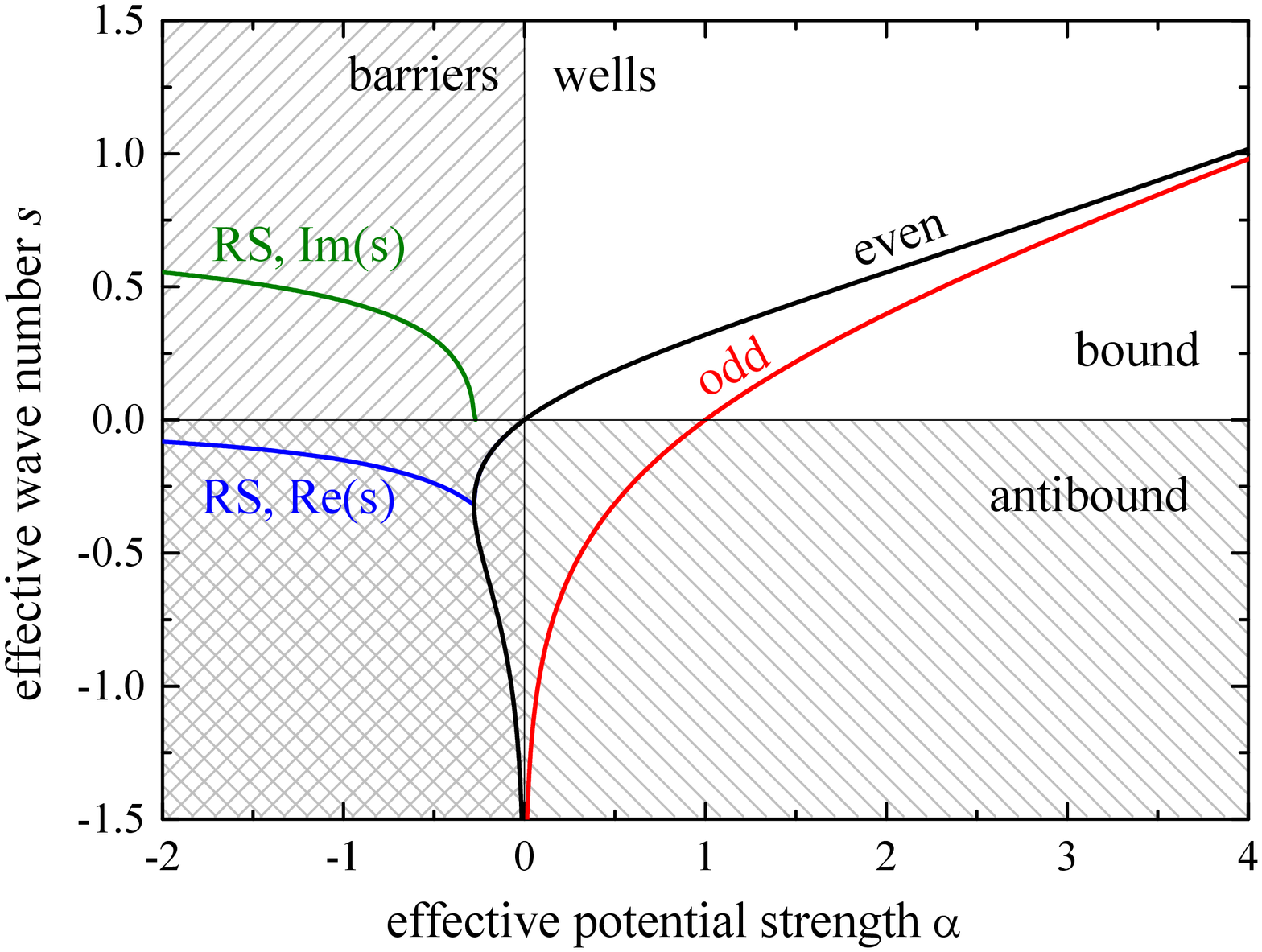}
\vskip-2cm
\caption{Effective wave number $s= -2ika$ as function of the effective potential strength $\alpha=\gamma a$ plotted for even and odd eigen states using the explicit functional dependence \Eq{sol-double-s}. The plot demonstrates transitions from bound to anti-bound states and then to normal RSs. Regions for potential wells ($\alpha>0$) and barriers ($\alpha<0$) are indicated, as well as for bound ($s>0$) and anti-bound ($s<0$) states. Blue and green lines show the real and imaginary parts of the wave numbers $s$ for the pair of the lowest energy normal RSs formed from a pair of degenerate anti-bound states at $\alpha\approx -0.27$. }
\label{fig:Double-s}
\end{figure}

To see it more clearly and also to investigate the dependence on the potential strength (e.g. keeping the width $2a$ fixed), we introduce another dimensionless imaginary wave number
$s=2\varkappa a$, so that \Eq{sol-double} takes the form
\be
\alpha=\frac{s}{1\pm e^{-s}}\,.
\label{sol-double-s}
\ee
The function $s(\alpha)$ can be easily plotted without solving the transcendental equation, due to the explicit functional dependence $\alpha(s)$ given by \Eq{sol-double-s}. Its plot is presented in \Fig{fig:Double-s}.
Since $a>0$, the region of positive $s$ corresponds to bound states. We see two bound states for $\alpha>1$ and only one for $0<\alpha<1$. The odd state transforms at $\alpha=1$  from bound to anti-bound, as negative $s$ corresponds to growing exponentials outside the systems, see \Eq{wf-double} for $k=is/2a$ and $s<0$. Another anti-bound state forms from the even bound state at $\alpha=0$ when the wells switch into barriers. At the same time, as $\alpha$ changes its sign from positive to negative, the odd  anti-bound state goes away to infinity as $s\to -\infty$ and then comes back at small negative values of $\alpha$ as an even anti-bound state, coexisting with the other even anti-bound state up to $\alpha\approx-0.27$. At that point the two anti-bound states merge, now transforming into a pair of normal RSs, which then evolve as $\alpha$ decreases further. Both RSs have the same imaginary part and the opposite real part of $k$, shown in \Fig{fig:Double-s} by blue and greens lines, respectively.

\subsection{Resonant states}
\label{sec-RSs}

\begin{figure}[t]
\includegraphics[scale=0.35,angle=-90]{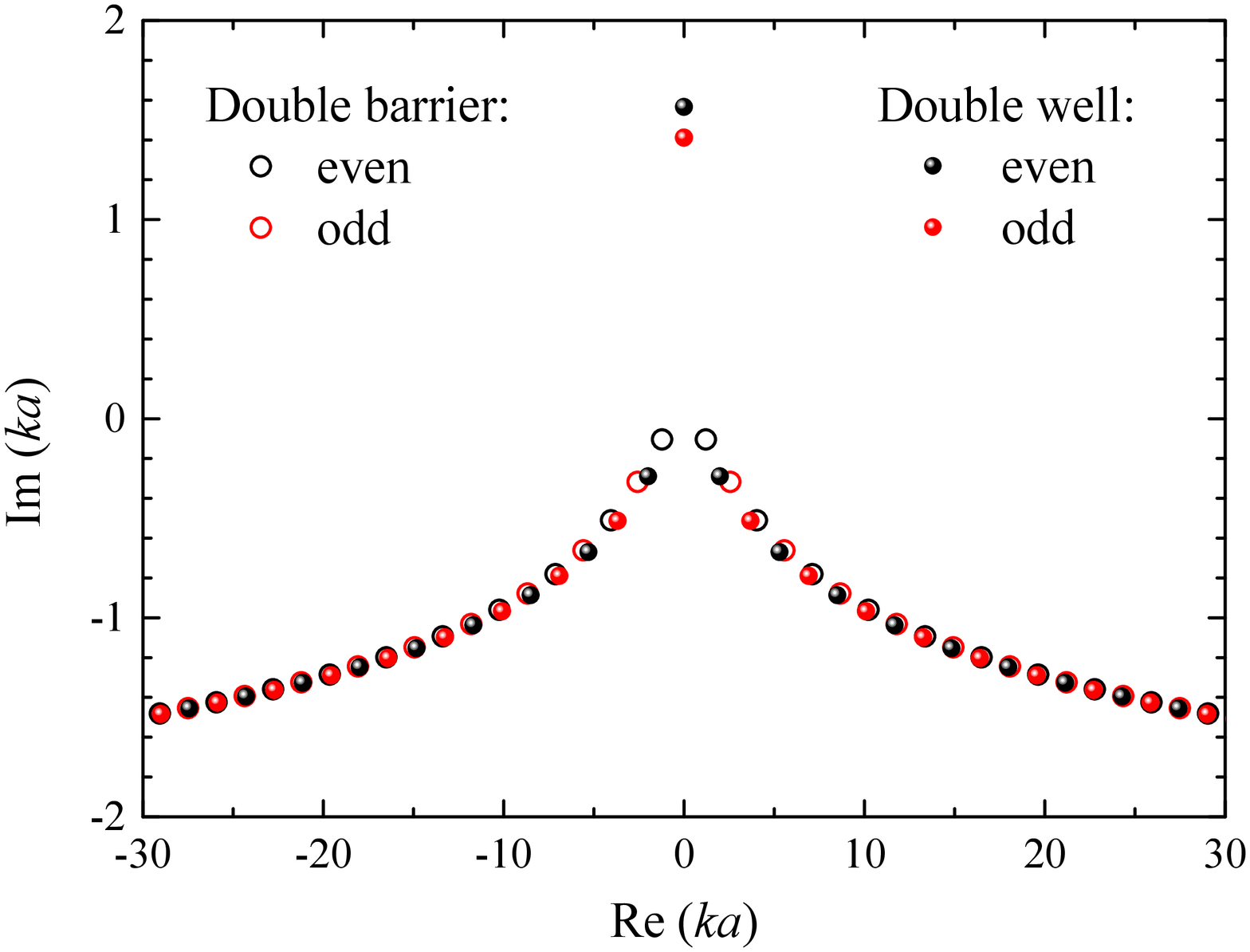}
\vskip-2cm
\caption{Complex eigen wave numbers $k_n$ of the RSs in a double delta-potential structure, with two wells ($\gamma a=3$, shiny balls) and two barrier ($\gamma a=-3$, open circles). Even and odd parity states are shown, respectively, in black and red.}
\label{fig:RSdouble}
\end{figure}

We now consider all possible solutions of \Eq{sol-double} in the complex $k$-plane, generating bound, anti-bound and normal RSs, as shown in \Fig{fig:RSdouble} for the case of a double well and a double barrier structure. For the wells ($\gamma a =3$), one can see two bound states and an infinite countable number of normal RSs with nonzero real and imaginary parts of $k$. Furthermore, these normal RSs all have complex wave functions which cannot be made real by redefining the normalization constant, unlike bound or anti-bound states. These RSs appear in pairs: Each RS with the eigen wave number $k$ and the wave function $\psi$ has a counterpart with the eigen wave number $-k^\ast$ and the wave function $\psi^\ast$, so that the spectra of RSs shown in \Fig{fig:RSdouble} possess a mirror symmetry about the imaginary axis, which is a general property of an open system, not related to its spatial symmetry. For the barriers ($\gamma a =-3$), there are only normal RSs seen in the spectrum, as this potential strength is too big for anti-bound states to exist, see \Fig{fig:Double-s}. In both spectra, normal RSs of even and odd parity appear in alternating order and are almost equally spaced for large $k$ dominated by the real part. The reason for that is that these states have the same nature as Fabry-P\'erot modes in an optical system, with a half integer multiple of their De-Broglie wavelength $\lambda=2\pi/k$ approximately matching the system size $2a$ Indeed, the spacing between the wave numbers plotted in \Fig{fig:RSdouble} is $\delta k\approx \pi/2a$. These RSs are formed from a constructive interference of waves created by multiple reflection from the potential inhomogeneities (wells or barriers) at $x=\pm a$.  The absolute value of the imaginary part of $k$ grows monotonously with the real part of $k$ (and consequently with the resonance energy), showing an increasing probability of a particle to leave the system as its energy increases.

Interestingly, at large $k$ the even RS wave numbers of the double well structure approach asymptotically the odd RS wave numbers of the double barrier structure, and vice versa, provided that the absolute values of $|\gamma a|$ are the same for the barriers and wells. This can be understood, looking again at \Eq{sol-double} and noticing that if
the first term was neglected, \Eq{sol-double} would become invariant with respect to a simultaneous flip of the sign of $\gamma$ (switching between barriers and wells) and the sign standing at the exponential (switching between even and odd solutions), thus making the above two cases equivalent. Indeed, this equivalence is asymptotically achieved at large $k$, when the first term in \Eq{sol-double} is getting small compared to the other two and can thus be neglected.

Applying the normalization condition \Eq{norm} to the wave function \Eq{wf-double} and excluding exponentials with the help of the secular equation \Eq{sol-double}, we find the normalization constants in \Eq{wf-double}:
\be
A=C\left(1+\frac{\gamma}{2ik}\right)^{-1},\ \ \    C=\frac{1}{2\sqrt{\pm[a-(\gamma+2ik)^{-1}]}}\,.
\label{norm-double}
\ee
The normalized wave functions of a double well or a double barrier system are now ready to use in the RSE which can be applied for various perturbations. This is however outside the scope of the present paper and will be presented elsewhere.

\section{Triple well}
\label{sec:triple}

\begin{figure}[t]
\vskip0.3cm
\hskip0.5cm
\includegraphics[scale=0.28,angle=-90]{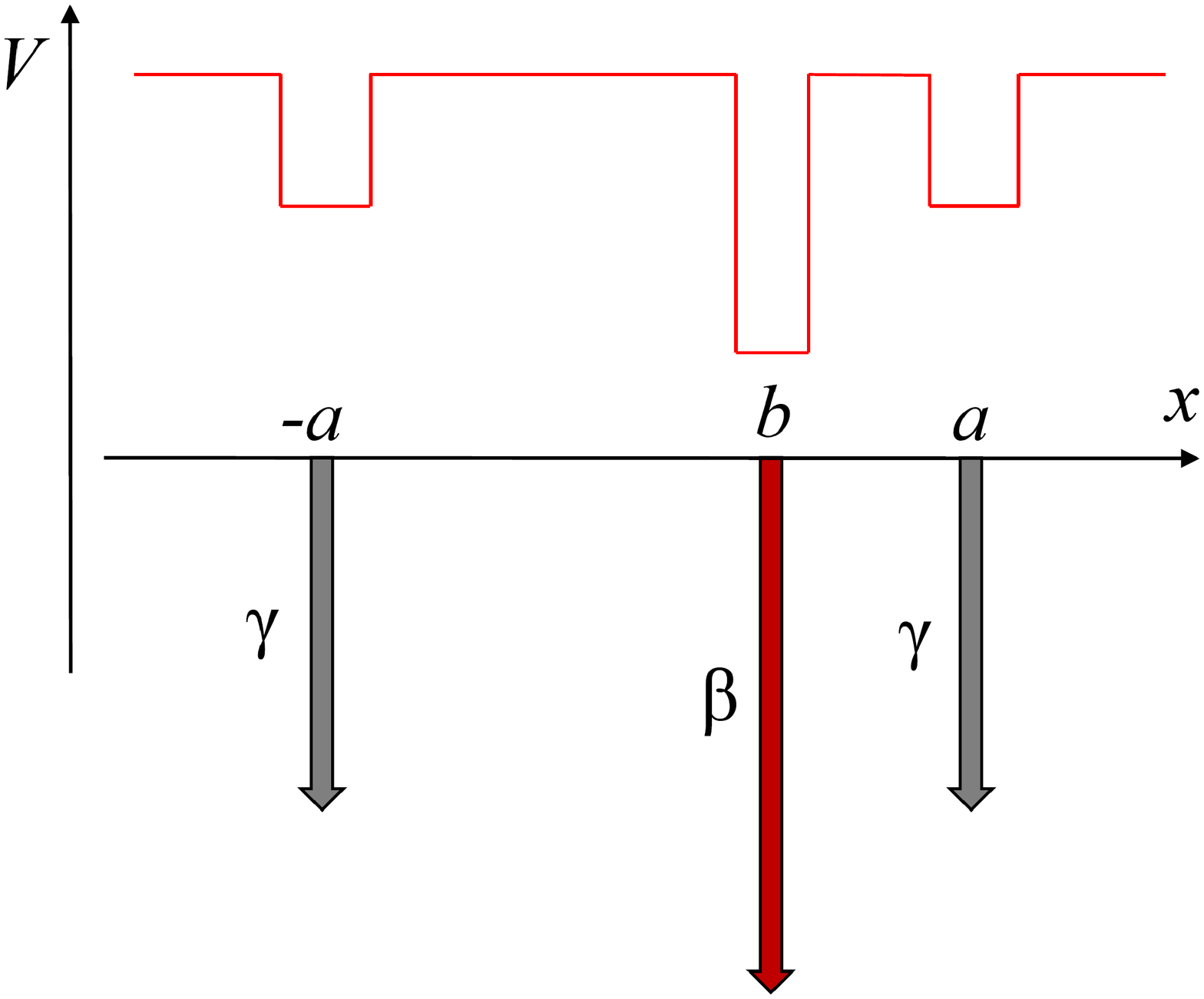}
%\vskip-2cm
\vskip-1.5cm
\caption{As \Fig{fig:Sketch-double} but for a triple well potential.}
\label{fig:Sketch-triple}
\end{figure}

We now add a third well (barrier) positioned at $x=b$, somewhere between the two equal wells (barriers): $-a<b<a$. It is modeled in the same way at the other two, so that the potential is given by
\be
V(x)=-\gamma\delta(x-a)-\gamma\delta(x+a)-\beta\delta(x-b)\,,
\label{pot-triple}
\ee
where the strength $\beta$ is generally different from $\gamma$, with $\beta>0$ ($\beta<0$) corresponding to an additional well (barrier). A sketch of this potential and its relation to a more realistic semiconductor heterostructure is provided in \Fig{fig:Sketch-triple}.

We use the same approach as in \Sec{sec:double} to solve the Schr\"odinger equation (\ref{SE}) with the potential \Eq{pot-triple}, taking the wave function of a RS in the following general form:
\begin{equation}
\psi(x) = \begin{cases}
A e^{ikx}                                &   x> a,\\
C_1e^{ikx}+ C_2e^{-ikx}           & b\leqslant x\leqslant a,\\
D_1e^{ikx}+ D_2e^{-ikx}           & -a\leqslant x<b,\\
B e^{-ikx}                                 &   x<-a.\\
\end{cases}
\label{wf-general}
\end{equation}

\subsection{Exact solution for a symmetric structure}

We first consider the case of a symmetric potential, having $b=0$ and $\beta$ arbitrary. Then using \Eq{even-odd}, we find $B=\pm A$ for the solution outside the system and
\be
C_1e^{ikx}+ C_2e^{-ikx}=D_1e^{-ikx}+ D_2e^{ikx}
\label{coef}
\ee
for the region inside it. Equating coefficients at the same exponentials in \Eq{coef}, obtain $D_2=\pm C_1= \pm C$ and $C_2=\pm D_1= \pm D$, where we have introduced constants $C$ and $D$ for brevity of notations. Then the wave function takes a simplified form:
\begin{equation}
\psi(x) = \begin{cases}
A e^{ikx}                                &   x> a,\\
C e^{ikx}\pm D e^{-ikx}         & 0\leqslant x\leqslant a,\\
D e^{ikx}\pm Ce^{-ikx}           & -a\leqslant x<0,\\
\pm A e^{-ikx}                                 &   x<-a.\\
\end{cases}
\label{wf-triple}
\end{equation}
The existence of the third delta function in the potential \Eq{pot-triple} leads to a new break in the derivative of the wave function at $x=0$, and to two more BCs:
\bea
\psi'(\varepsilon)-\psi'(-\varepsilon)&=&-\beta\psi(0)\,,
\label{BC3}
\\
\psi(+\varepsilon)-\psi(-\varepsilon)&=&0\,,
\label{BC4}
\eea
in addition to the pair of BCs given by Eqs.\,(\ref{BC1}) and (\ref{BC2}). Using \Eq{BC4} for an odd parity state [the lower sign in \Eq{wf-triple}] results in a condition $C=D$ meaning that $\psi(0)=0$, as should be for any anti-symmetric state. This makes however the odd state insensitive to the potential well or barrier if the latter is placed exactly in the center of the system, thus keeping $\psi'(x)$ continuous at $x=0$. The odd parity solution of the Schr\"odinger equation with the potential \Eq{pot-triple} and $b=0$ is thus the same as for the double delta potential \Eq{pot-double} and is described in detail in \Sec{sec:double}. We therefore concentrate below on even parity states.

For even parity states, \Eq{BC4} is automatically fulfilled due to \Eq{even-odd}, but \Eq{BC3} brings in a unique information about the middle well/barrier: $2ik(C-D)=-\beta(C+D)$, or
\be
\sigma=\frac{D}{C}=-\frac{1+2ik/\beta}{1-2ik/\beta}\,.
\label{DC}
\ee
At the same time, Eqs.\,(\ref{BC1}) and (\ref{BC2}) now give
\bea
ik A e^{ik a}-ik (C e^{ika}-D e^{-ika})&=&-\gamma A e^{ik a}\,,
\label{BC1b}
\\
A e^{ik a}-(C e^{ika}+D e^{-ika})&=&0\,,
\label{BC2b}
\eea
which result, after having combined them with \Eq{DC}, in two different expressions for the ratio $A/C$,
\be
\frac{A}{C}=\frac{ik (e^{ika}-\sigma e^{-ika})}{(ik+\gamma)e^{ika}}=\frac{e^{ika}+\sigma e^{-ika}}{e^{ika}}\,,
\ee
determining the secular transcendental equation for even-parity states:
\be
1+\frac{2ik}{\gamma}= \frac{1-2ik/\beta}{1+2ik/\beta} e^{2ika}\,.
\label{sol-triple}
\ee
In the limit $\beta\to0$, \Eq{sol-triple} reduces back to the secular equation (\ref{sol-double}) for the even-parity (ground) state of the double quantum well.

\subsection{Bound and anti-bound states}
\label{sec:triple-bound}

\begin{figure}[t]
\includegraphics[scale=0.35,angle=-90]{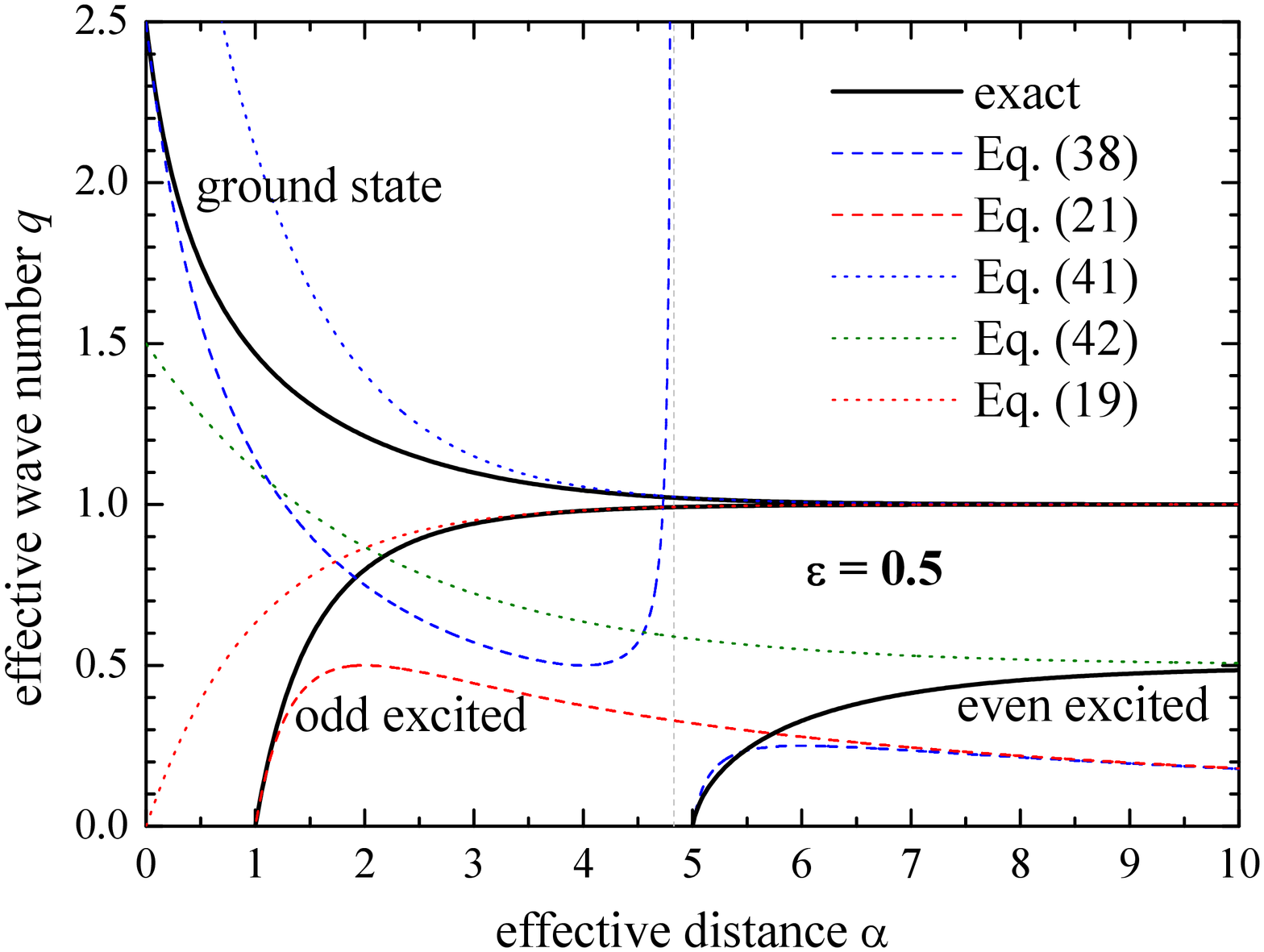}
\vskip-2cm
\caption{As \Fig{fig:Double-q} but for a triple quantum well system modeled by a triple-delta potential with $b=0$ and $\varepsilon=1/2$.}
\label{fig:Triple-q}
\end{figure}

Repeating the procedure used in \Sec{sec:double-bound}, we first introduce a purely imaginary wave number $k=i\varkappa$, expressed in terms of a real valued $\varkappa$ and then use an effective dimensionless wave number $q=2\varkappa/\gamma$, in order to study the dependence of the bound state on the system size. In addition to the effective width/strength $\alpha$ defined by \Eq{alpha}, we introduce a relative strength of the middle well/barrier:
\be
\varepsilon=\frac{\beta}{\gamma}\,.
\label{eps-def}
\ee
Equation (\ref{sol-triple}) then takes the form
\be
q=1+\frac{q+\varepsilon}{q-\varepsilon}e^{-q\alpha}
\label{sol-triple-q}
\ee
[compare with \Eq{sol-double-q} for $+$]. The exact numerical solution of \Eq{sol-triple-q} for $\varepsilon=0.5$ is shown in \Fig{fig:Triple-q} by black solid lines displaying two even parity bound states, as well as the odd parity state which is the same as in \Fig{fig:Double-q}. While the ground state having the highest value of $q$ exists for any size of the system (i.e. for all $\alpha>0$), the 2nd excited (even) state disappears in this case at $\alpha=5$.

To understand this behavior, we again use the Taylor expansion \Eq{Taylor} up to 2nd order, obtaining from \Eq{sol-triple-q} an approximation for even states:
\be
q\approx\frac{2-\varepsilon(\alpha-1)}{1+\alpha(1-\alpha\varepsilon/2)}\,.
\label{triple-approx}
\ee
Obviously, for $\varepsilon=0$, \Eq{triple-approx} is equivalent to the approximation \Eq{double-approx} for the even parity values $q_+$. The approximation \Eq{triple-approx} is shown for $\varepsilon=0.5$ in \Fig{fig:Triple-q} by dashed blue lines, demonstrating a good agreement with the full solution for $\alpha\to 0$ (ground state) and for $q\to0$ (2nd excited states). The last limit allows us to obtain the following inequality for $\varepsilon$ and $\alpha$:
\be
\alpha>1+\frac{2}{\varepsilon}\,,
\label{even-inequ}
\ee
showing under which conditions an even bound state exists.
When both $\gamma>0$ and $\beta>0$, this inequality refers to the 2nd excited state in a triple well. In particular, for the example in \Fig{fig:Triple-q}, the even excited state disappears at $\alpha=1+2/\varepsilon =5$.

\begin{figure}[t]
\includegraphics[scale=0.35,angle=-90]{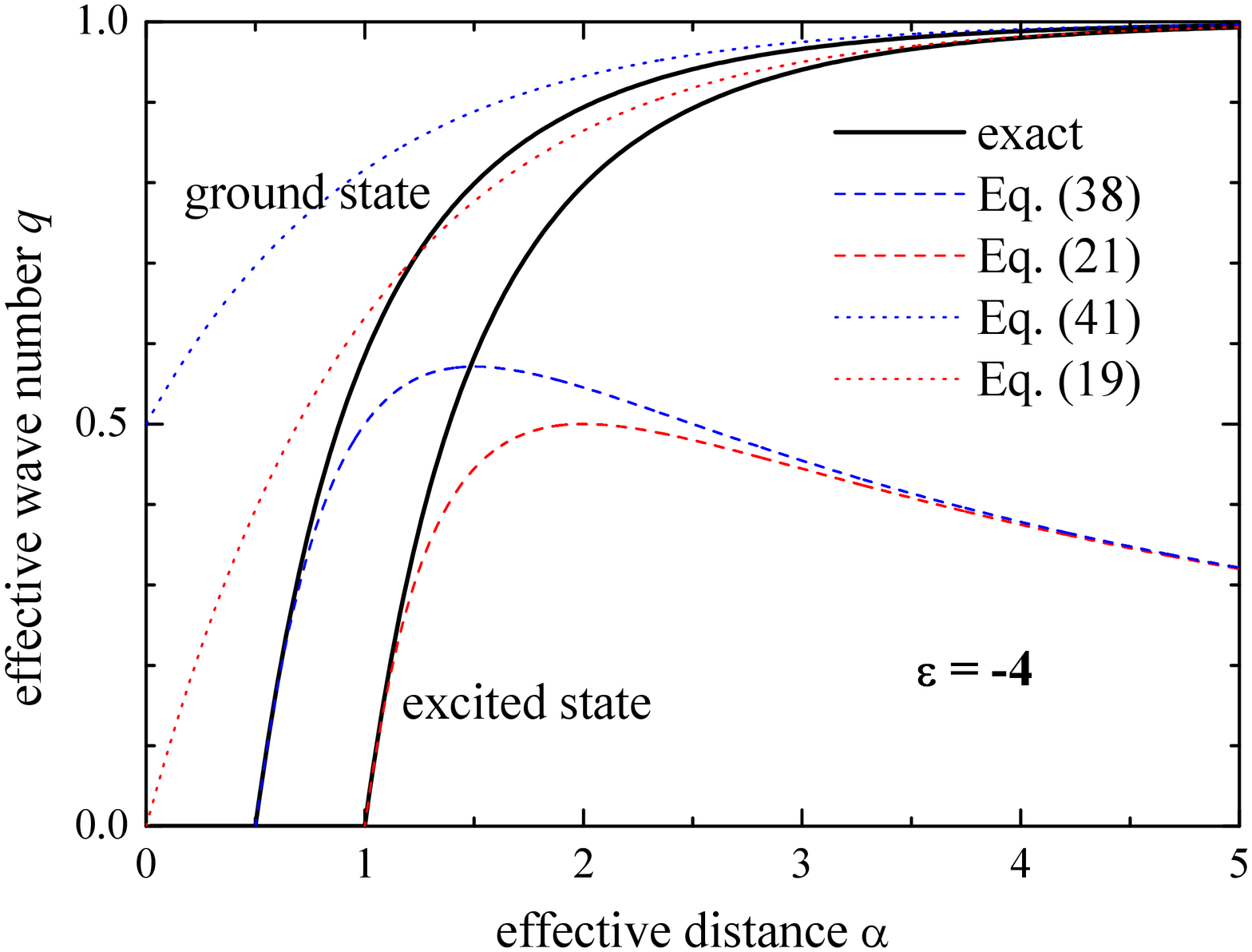}
\vskip-2cm
\caption{As \Fig{fig:Triple-q} but for $\varepsilon=-4$.  }
\label{fig:Triple-q2}
\end{figure}

If, however, there is a barrier in the middle of two wells, i.e. $\gamma>0$ but $\beta<0$, there is a maximum of two bound states in the spectrum, one even (the ground state) and one odd (the excited state), and the same \Eq{even-inequ} now becomes a condition for the ground state to exist. Indeed, if the barrier is high enough, namely if $\beta<-2\gamma$, the ground state also disappears from the spectrum at the system size  smaller than that determined by \Eq{even-inequ}. This case presents an interesting situation when a one-dimensional symmetric potential well structure cannot accommodate any bound states. An illustration for $\varepsilon=-4$ is provided in \Fig{fig:Triple-q2} showing that the ground state disappears at $\alpha=1/2$, in agreement with \Eq{even-inequ} and \Fig{fig:Triple-eps} below.

To analyze the behavior at large system sizes, we take the limit $\alpha\to\infty$, which makes the exponential term in \Eq{sol-triple-q} small. This results in a quadratic equation for $q$:
\be
q^2-q(1+\varepsilon+e^{-q\alpha})+\varepsilon(1-e^{-q\alpha})=0
\ee
giving solutions
\be
q_0\approx 1+\frac{1+\varepsilon}{1-\varepsilon}e^{-\alpha}
\ee
for the ground and
\be
q_2\approx \varepsilon+\frac{\varepsilon}{1-\varepsilon}e^{-\varepsilon\alpha}
\ee
for the 2nd excited state. These approximate values are also plotted in Figs.\,\ref{fig:Triple-q} and \ref{fig:Triple-q2} showing an agreement with the full solution.

\begin{figure}[t]
\includegraphics[scale=0.35,angle=-90]{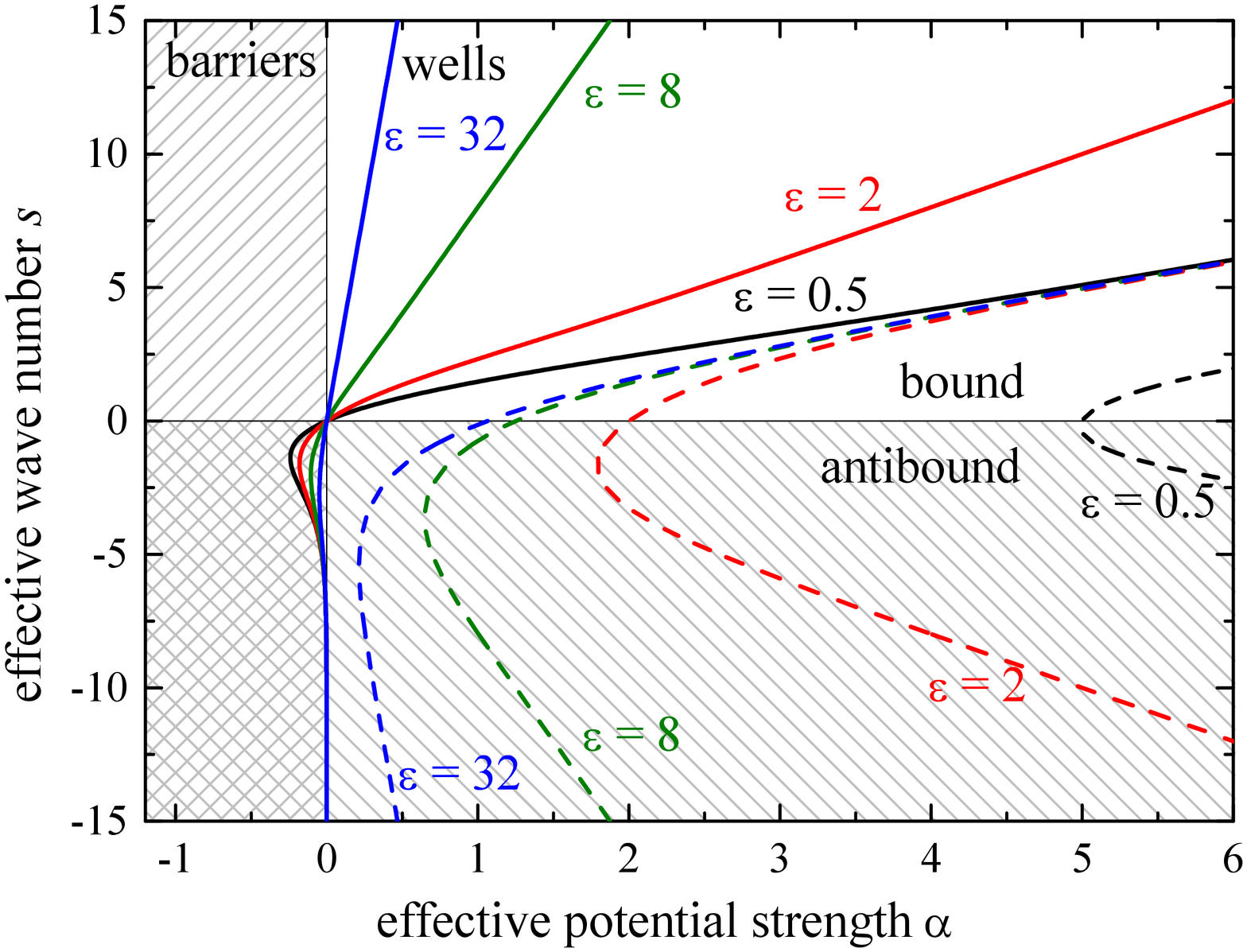}
\vskip-2cm
\caption{As \Fig{fig:Double-s} but for a triple-delta potential with $b=0$ and different $\varepsilon$ as labeled. All curves are obtained by plotting the inverse of the function \Eq{sol-triple-s}. Note that only even parity solutions are displayed, with an excited state branch shown by a dashed curve for each $\varepsilon$.}
\label{fig:Triple-s}
\end{figure}

To study the dependence on the quantum well strength $\gamma$ (i.e. $\alpha$ for a fixed $a$), we introduce, as in \Sec{sec:double-bound}, the effective wave number $s=2\varkappa a$. Then \Eq{sol-triple} becomes
\be
\frac{s}{\alpha}=1+\frac{s/\alpha+\varepsilon}{s/\alpha-\varepsilon} e^{-s}
\ee
which has an explicit solution for $\alpha(s)$:
\be
\alpha=\frac{2s}{1+\varepsilon+e^{-s}\pm \sqrt{(1-\varepsilon)^2+2(1+3\varepsilon)e^{-s}+e^{-2s}}}\,,
\label{sol-triple-s}
\ee
[compare with \Eq{sol-double-s}]. Again, the advantage of representing the solution in the form of \Eq{sol-triple-s} is that it can be displayed without solving the secular equation. The plots of it are presented in \Fig{fig:Triple-s}, showing the evolution of bound and anti-bound states with the effective potential strength $\alpha$. We see that as $\alpha$ decreases  the bound states transform into anti-bound states and then to normal RSs (not shown in \Fig{fig:Triple-s}), as in the case of a double barrier, see \Fig{fig:Double-s}.

Finally, by fixing $\alpha$ (i.e. the product of the potential strength $\gamma$ and the width $a$) the dependence $q(\varepsilon)$ or $s(\varepsilon)$ on the relative potential strength $\varepsilon$, given by \Eq{eps-def}, can be extracted. Expressing $\varepsilon$ from \Eq{sol-triple-q} obtain
\be
\varepsilon=q\,\frac{1-q+e^{-q\alpha}}{1-q-e^{-q\alpha}}=\frac{s}{\alpha}\,\frac{1-s/\alpha+e^{-s}}{1-s/\alpha-e^{-s}}\,.
\label{epsilon}
\ee
Taking the inverse of this function, we find the dependence $s(\varepsilon)$ [or $q(\varepsilon)$] which is displayed in \Fig{fig:Triple-eps}, showing the evolution of states with the the potential ratio $\varepsilon$ continuously changing between positive and negative values, thus covering also an important case of mixed potentials (with a barrier in the middle). Interestingly the two even states displayed show a sort of avoided crossing which is getting sharper with increased potential strength/width $\alpha$, owing to a smaller tunnel coupling between the wells.
\begin{figure}[t]
\includegraphics[scale=0.35,angle=-90]{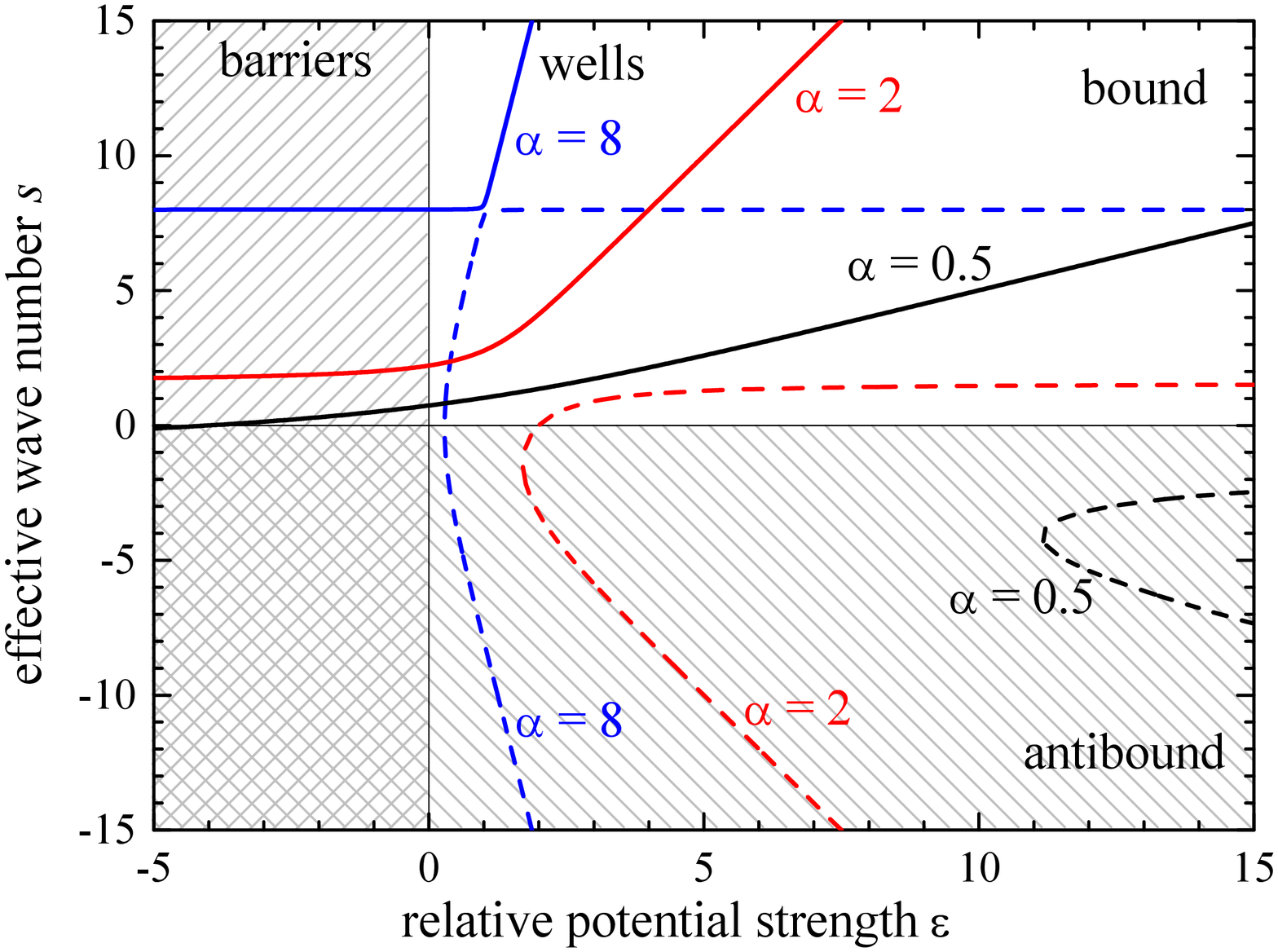}
\vskip-2cm
\caption{As \Fig{fig:Triple-s} but for the relative potential strength dependence given by \Eq{epsilon}, for different values of $\alpha$ as labeled.}
\label{fig:Triple-eps}
\end{figure}

\subsection{Solutions for an asymmetric case}

Let us now consider a more general case of an arbitrary position of the middle well/barrier at $x=b$, with $-a<b<a$. To find the secular equation for RSs and relations between 6 amplitudes in the wave function \Eq{wf-general}, one needs to satisfy 3 pairs of BCs, describing the continuity of the wave function and discontinuity of its first derivative at $x=-a$, $b$ and $a$. We skip details of this derivation, which can be made in a similar way to Secs.\ref{sec:double-bound} and \ref{sec:triple-bound}. We present  a resulting secular equation for $k$, which can be written compactly as
\be
\xi^2 (1-\eta) - 2\xi \cos (2kb) +1+\eta=0\,,
\label{triple}
\ee
after introducing
\be
\xi=\frac{e^{2ika}}{1+2ik/\gamma}\ \ \ {\rm and}\ \ \ \eta=\frac{2ik}{\beta}\,.
\label{xi1}
\ee

We again first study the dependence of the full solution of \Eq{triple} for bound states on the potential strength $\gamma$, for fixed $a$, $b$ and $\beta$. Introducing $\alpha=\gamma a$ and $s=-2ika$, as before, and solving the quadratic equation (\ref{triple}) for $\xi$, obtain two branches of the solution:
\be
\alpha_\pm(s)=\frac{s}{1-e^{-s}/\xi_\pm(s)}\,,
\label{alpha-pm}
\ee
where
\be
\xi_\pm(s)=\frac{c\pm\sqrt{c^2+\eta^2-1}}{1-\eta}\,,
\label{xi}
\ee
\be
c(s)=\cosh(sb/a)\ \ \ {\rm and } \ \ \ \eta(s)=-\frac{s}{\beta a}\,.
\ee
It is instructive to see that for $b=0$, the two branches become
\be
\xi_-=1\ \ \ {\rm and}\ \ \ \xi_+=\frac{1+\eta}{1-\eta}
\ee
corresponding to the odd and even parity states and coinciding with the solution for a symmetric triple well structure given by \Eq{sol-double} with the lower sign used and by \Eq{sol-triple}, respectively. Taking further the limit $\eta\to\infty$ (corresponding to $\beta\to0$), obtain solutions for a double well structure: $\xi_\mp = \pm 1$ which, after substitution into \Eq{xi}, give exactly \Eq{sol-double-s}.

Finally, expressing $\eta$ from \Eq{triple}, we find an explicit dependence of the wave numbers $k$ on the middle well/barrier strength $\beta$:
\be
\beta a = s\,\frac{1-\xi^2}{1-2\xi c +\xi^2}\,,
\label{beta}
\ee
where  $\xi= e^{-s}/(1-s/\alpha)$, in accordance with \Eq{xi1}. The function $k(\beta)$ can be obtained by simply inverting the function $\beta(k)$ given by \Eq{beta}.
For the symmetric quantum well structure ($b=0$), using $c=1$ in  \Eq{beta} leads to
\be
\beta a = s\,\frac{1+\xi}{1-\xi}\,,
\ee
\begin{figure}[t]
%\vskip-1cm
\includegraphics[scale=0.35,angle=-90]{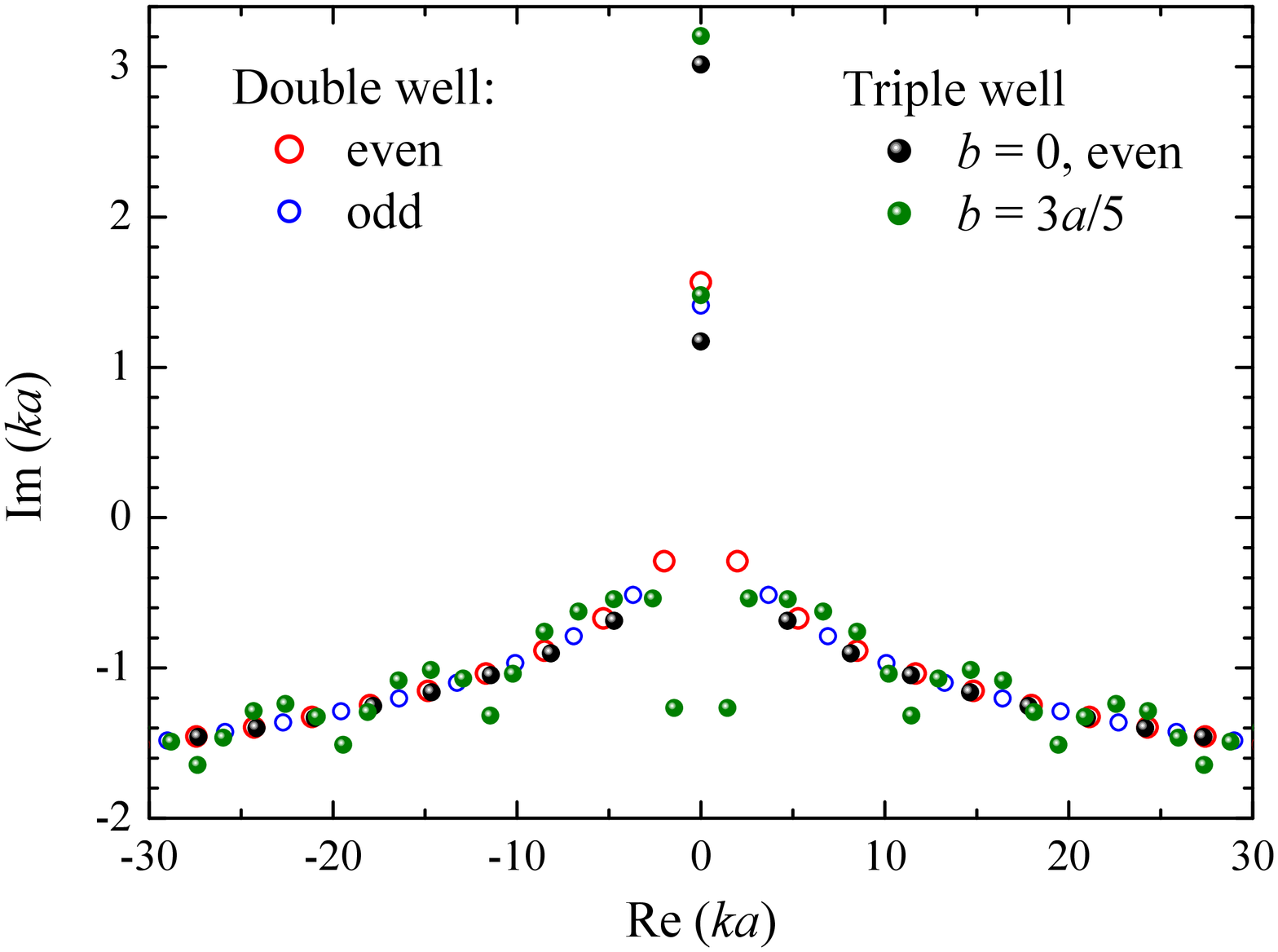}
\vskip-1cm
\caption{As \Fig{fig:RSdouble}, for double ($\gamma=3/a$), triple symmetric ($b=0$, $\gamma=3/a$, $\beta=6/a$)  and triple asymmetric ($b=3a/5$, $\gamma=3/a$, $\beta=6/a$) quantum well structures.}
\label{fig:RStriple}
\end{figure}
which is exactly the same as \Eq{epsilon}.

\subsection{Resonant states}

RSs for both symmetric and asymmetric triple well structures are shown in \Fig{fig:RStriple}. The spectrum of RSs for the symmetric structure  is quite similar to that of the double well which is also shown for comparison (the same as demonstrated in \Fig{fig:RSdouble}). Note that odd RSs remain the same for both systems. For the triple well, we now see two bound states, in accordance with our analysis in \Sec{sec:triple-bound}. Indeed, for $\alpha=3$ and $\varepsilon=2$ the inequality Eq.\,(\ref{even-inequ}) is fulfilled allowing the second excited state to exist.

The spectrum of RSs for an asymmetric triple well structure with $b=3a/5$, $\gamma=3/a$ and $\beta=6/a$ is quite different. First of all, being shifted from the center of the structure, the middle quantum well mixes even and odd RSs. As a result, stronger deviations from the double well spectrum of RSs is seen. Choosing the ratio $2a/(a-b)$ equal to an integer, as in the present case, the third well in the middle splits the structure into two resonators having commensurable widths $a_L$ and $a_R$. In our case, $a_L=8a/5$ and $a_R=2a/5$, thus splitting the full width of the system in 4:1 proportion. Therefore resonances accommodated in the right (narrower) subsystem can be enhanced, owing to an additional constructive interference of wave, by the left (wider) subsystem and the full-width structure. As a result, one can see a quasi-periodic modulation in the spectrum with the period of about $\pi/a_R$, which is five times larger than the separation between the RS wave numbers, which is approximately $\pi/(2a)$, see \Sec{sec-RSs}.

\section{Role of the resonant states in the transmission}

In this section, we study the role of RSs in observables, such as the local density of states or the scattering matrix. Below we consider, as an example, the transmission of a particle  through a quantum system consisting of two Dirac delta wells. A particle traveling in free space is described by a wave function in the form of plane wave with a wave number $k$. We first calculate analytically its transmission amplitude $t_a(k)$ as a function of the {\it real} wave number $k$ of the particle. This transmission can be found by choosing appropriate BCs outside the system, namely by allowing the system to be excited with an incoming wave. To do so, we keep in \Eq{psi0} the term with an incoming wave which now has a non-vanishing amplitude $D\neq0$, while requiring that $B=0$. The BCs at $x=\pm a$, given by Eqs.\,(\ref{BC1}) and (\ref{BC2}), are the same as for the RSs. Applying them and solving a set of algebraic equations, we find the transmission amplitude
\be
t_a(k)=\frac{A}{D}=\frac{4k^2}{4k(k-i\gamma)-\gamma^2(1-e^{4ika})}\,.
\label{ta}
\ee
\begin{figure}[t]
\vskip-0.5cm
\includegraphics[scale=0.35,angle=-90]{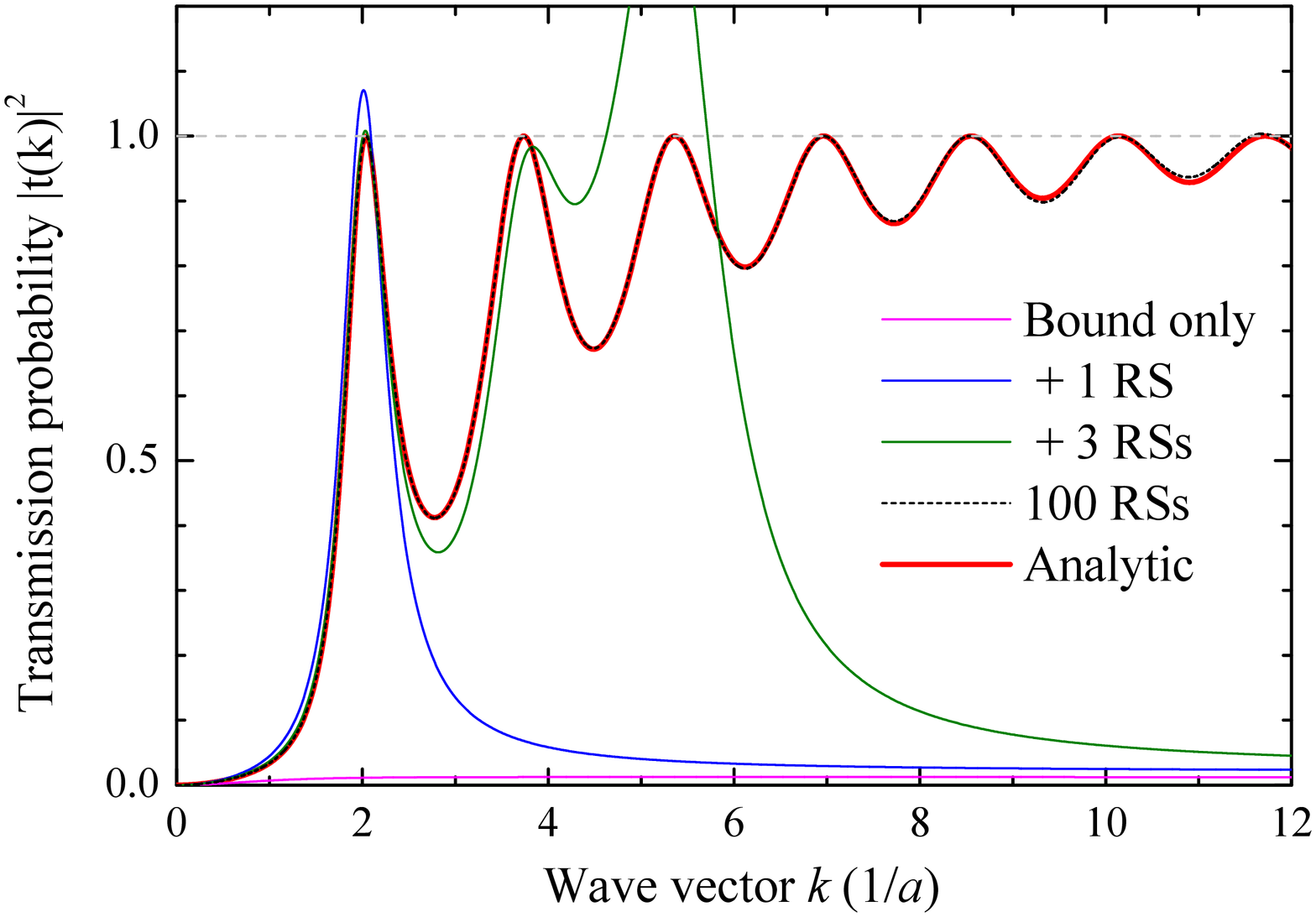}
\vskip-0.5cm
\caption{ The probability of transmission trough a double Dirac delta well structure with $\gamma=3/a$, evaluated using the analytic transmission $t_a(k)$ (thick red curve)  and its spectral representation $t(k)$ calculated for different number of RSs taken into account in the Mittag-Leffler expansion: bound states only (magenta), bound states and 1 pair of RSs (dark blue), bound states and 3 pairs of RSs (dark green), and 100 RSs in total, including the bound states (black dashed curve).}
\label{fig:Trans}
\end{figure}
Now taking the analytic continuation of this function into the complex $k$ plane, it is easy to see that $t_a(k)$ has simple poles at $k=k_n$, where $k_n$ are the wave numbers of all possible RSs (including bound, anti-bound and normal RSs), which satisfy the secular \Eq{sol-double}.

For an {\it arbitrary} one-dimensional potential with compact support, i.e. vanishing (or constant) outside the system area $|x|\leqslant a$, the transmission amplitude is given by
\be
t(k)=2ik e^{-2ika} G_k(a,-a)\,,
\label{tG}
\ee
see e.g.~\cite{DoostPRA12}. Here, $G_k(x,x')$ is the Green's function of the Schr\"odinger equation~(\ref{SE}) for a given fixed wave number $k$. For the coordinates $x$ and $x'$ within the system, the Green's function is vanishing on an infinitely large circle in the complex $k$ plane, and therefore,  one can apply to it the Mittag-Leffler (ML) theorem which yields~\cite{MorePRA71,MuljarovEPL10,ArmitagePRA14}
\be
G_k(x,x')=\sum_n\frac{\psi_n(x)\psi_n(x')}{2k_n(k-k_n)}\,,
\label{GF-ML}
\ee
where $\psi_n(x)$ are the RS wave functions normalized according to \Eq{norm}.

For illustration, we apply the general result given by  Eqs.\,(\ref{tG}) and (\ref{GF-ML}) to the particular case of the double delta-function potential \Eq{pot-double}. Using the explicit form of the wave functions \Eq{wf-double}, their normalization \Eq{norm-double}, and the secular equation (\ref{sol-double}), one can write the transmission, with the help of Eqs.\,(\ref{tG}) and (\ref{GF-ML}), in the form of an infinite series over its poles:
\be
t(k)= k e^{-2ika}\sum_{n}\frac{R_n}{k-k_n}\,,
\label{ML}
\ee
where
\be
R_n= \pm\frac{i k_n a^2}{\gamma[(\gamma+2ik_n)a-1]}\,.
\label{Rn}
\ee
This result can be compared with the analytic transmission $t_a(k)$, given by \Eq{ta}, which is done in \Fig{fig:Trans}.

Using the ML expansion Eqs.\,(\ref{ML}) and (\ref{Rn}), we also study in \Fig{fig:Trans} the role of different RSs in the transmission. We first note that in this representation, bound states play a small but non-negligible role, producing some background contribution. The maxima of the transmission reaching the value of 1 for this symmetric quantum structure can be described by only taking into account in the summation  \Eq{ML} the corresponding normal RSs. Adding the very first pair of normal RSs already describes quite well the first peak in the transmission. The agreement is further improved by adding more RSs. With three pairs of RSs, the first peak of the transmission is fully reproduced, but the other two are described only qualitatively. To correct this and to describe other peaks, more RSs in \Eq{ML}  are needed. Taking all of them into account, the correct transmission is fully reproduced.

\section{Conclusion}

We have studied the full set of resonant states of a one-dimensional Schr\"odinger problem with double and triple quantum wells or barriers approximated by Dirac delta functions.
This full set includes bound, anti-bound and normal resonant states. We have revisited the problem of finding bound states in delta-well potentials and have worked out simple analytic expressions for important limiting cases and compared them with the full numerical solution. The latter is in turn presented here as universal dependencies containing the minimal number of parameters. Furthermore, we have studied the transition between different types of resonant states, demonstrating in particular how bound states disappear from the spectrum continuously transforming into anti-bound states, which are in turn transform further into normal resonant states. We have shown that these normal resonant states  determine the main spectral features in observables, such as the quantum transmission, and that taking the full set of resonant states, including the bound and anti-bound states,  allows one to precisely determine the transmission via its Mittag-Leffler expansion. Finally, we have analyzed the RSs of double and triple quantum wells in terms of the constructive interference of quantum waves supported by these structures.

\bibliography{RSs}

\end{document}